\documentclass[conference]{IEEEtran}

\usepackage{cite}
\usepackage{graphicx}
\usepackage{amsmath, amssymb}
\usepackage{algorithm}
\usepackage{algpseudocode}
\usepackage{booktabs}
\usepackage{multirow}
\usepackage{tabularx}

\begin{document}

\title{AI-Driven Dynamic Firewall Optimization Using Reinforcement Learning for Anomaly Detection and Prevention}

\author{
    \IEEEauthorblockN{Taimoor Ahmad}
    \IEEEauthorblockA{dept. of Computer Science \\
    The Superior Univeristy Lahore\\
Lahore, Pakistan \\
    Taimoor.ahmad1@superior.edu.pk}

}

\maketitle

\begin{abstract}
The growing complexity of cyber threats has rendered static firewalls increasingly ineffective for dynamic, real-time intrusion prevention. This paper proposes a novel AI-driven dynamic firewall optimization framework that leverages deep reinforcement learning (DRL) to autonomously adapt and update firewall rules in response to evolving network threats. Our system employs a Markov Decision Process (MDP) formulation, where the RL agent observes network states, detects anomalies using a hybrid LSTM-CNN model, and dynamically modifies firewall configurations to mitigate risks. We train and evaluate our framework on the NSL-KDD and CIC-IDS2017 datasets using a simulated software-defined network environment. Results demonstrate significant improvements in detection accuracy, false positive reduction, and rule update latency when compared to traditional signature- and behavior-based firewalls. The proposed method provides a scalable, autonomous solution for enhancing network resilience against complex attack vectors in both enterprise and critical infrastructure settings.
\end{abstract}

\section{Introduction}

The rapidly evolving landscape of cyber threats has outpaced the capabilities of traditional, static firewalls, which rely heavily on predefined rules and signature-based detection methods. As enterprise networks grow in scale and complexity, the ability to adapt to new threats in real time becomes a critical requirement. In this context, artificial intelligence (AI), and particularly deep reinforcement learning (DRL), offers a promising solution to automate firewall rule management and anomaly detection based on real-time network behavior\cite{z73,z74}.

A firewall is a fundamental security component that acts as a barrier between trusted and untrusted networks. Conventional firewalls apply rule-based packet filtering, where each incoming or outgoing packet is checked against static policies. Although effective for known attack vectors, these systems fall short when facing zero-day attacks, polymorphic malware, or distributed denial-of-service (DDoS) attacks~\cite{garcia2020survey,z71,z72}. Moreover, frequent manual reconfiguration increases administrative overhead and introduces human error.

Recent advancements in machine learning (ML) for cybersecurity have led to the development of anomaly-based intrusion detection systems (IDS) that learn patterns of normal behavior and flag deviations. However, these systems typically operate in a passive mode, alerting administrators but not taking active mitigation steps. Reinforcement learning (RL), with its ability to learn sequential decision-making tasks, provides a mechanism to actively intervene, adjust policies, and optimize defense strategies in a closed feedback loop~\cite{nguyen2019deep,z3333,z55}.

Our proposed solution, AI-Driven Dynamic Firewall Optimization (ADF-RL), builds a dynamic firewall system on top of a deep reinforcement learning agent that can intelligently update firewall rules in real time. The agent observes network traffic features and their temporal patterns via a hybrid LSTM-CNN anomaly detector and responds by inserting, removing, or reordering rules in the firewall based on the evolving threat landscape.

In ADF-RL, the environment is modeled as a Markov Decision Process (MDP), where states represent feature vectors of network traffic, actions correspond to rule changes, and rewards are determined by successful threat mitigation and low false-positive rates. Unlike traditional firewalls, our system can learn policies from interactions with traffic data, achieving continuous improvement and adaptability without human intervention~\cite{zhao2020rlfirewall,z5}.

Solving this problem is crucial for modern network infrastructure, particularly in high-stakes environments such as financial institutions, healthcare systems, and critical government services. Automating firewall management reduces response time, minimizes configuration errors, and strengthens resilience against unknown attacks.

\textbf{Key contributions of this paper are:}
\begin{itemize}
  \item We propose a novel deep reinforcement learning-based firewall architecture capable of dynamic rule adaptation in response to detected anomalies.
  \item We design a hybrid LSTM-CNN model for efficient anomaly detection that guides policy learning in the reinforcement agent.
  \item We implement the framework within a simulated SDN environment using NSL-KDD and CIC-IDS2017 datasets and compare it against static and ML-enhanced firewalls.
  \item We demonstrate improved detection accuracy, reduced false positive rates, and faster rule updates, validating the practical applicability of our approach.
\end{itemize}

The rest of this paper is organized as follows. Section II reviews related work on AI-based intrusion detection and adaptive firewall systems. Section III details the system model and mathematical formulation. Section IV describes the experimental setup and evaluates performance. Section V concludes the study and outlines future research directions.

\section{Related Work}

Several studies have explored the integration of machine learning and reinforcement learning in the context of intrusion detection and firewall management. This section reviews ten notable contributions in this domain and identifies gaps addressed by our proposed ADF-RL framework.

Zhang et al.~\cite{zhang2019anomaly} proposed a deep learning-based anomaly detection system using stacked autoencoders on the NSL-KDD dataset. While effective in identifying anomalies, their system lacked autonomous response mechanisms, relying solely on alerts for administrators.

Shen et al.~\cite{shen2020reinforcement} implemented a reinforcement learning-based firewall policy tuner for software-defined networks (SDNs). Though the system demonstrated adaptability, it did not incorporate deep feature extraction models, limiting its ability to detect stealthy threats.

Li et al.~\cite{li2021smartfire} introduced SmartFire, a reinforcement learning framework that dynamically adjusts firewall rule priorities. However, their method was limited to fixed rule sets and did not support real-time anomaly detection from live traffic data.

Wang et al.~\cite{wang2020dynamic} combined a decision tree classifier with Q-learning for adaptive firewall tuning. Their system struggled with high-dimensional feature spaces and generated numerous false positives under complex attack simulations.

Alshamrani et al.~\cite{alshamrani2020deep} developed a CNN-based intrusion detection engine for IoT environments. While their detection accuracy was high, the system required external control for policy enforcement and was not self-adaptive.

Kim et al.~\cite{kim2021selfadaptive} explored self-adaptive security agents using reinforcement learning. They successfully demonstrated policy learning under dynamic conditions but focused more on access control than fine-grained firewall rule management.

Gupta et al.~\cite{gupta2019towards} applied deep Q-networks (DQN) for IDS rule optimization. Although effective, their system lacked the ability to capture long-term temporal correlations in network behavior.

Tian et al.~\cite{tian2022lstm} proposed an LSTM-based IDS for SCADA systems, emphasizing time-sequence modeling. However, their model was primarily diagnostic and did not influence proactive firewall strategies.

Huang et al.~\cite{huang2019mlfirewall} built an ML-enhanced firewall with rule optimization based on supervised learning. Their static dataset assumption made it impractical for real-time deployment.

Rahman et al.~\cite{rahman2021cyberdefense} surveyed cyber-defense strategies combining anomaly detection and automated response but emphasized design-level insights rather than implementation or quantitative benchmarking.

In summary, existing solutions either rely heavily on static rule sets, lack real-time responsiveness, or use machine learning in a diagnostic rather than adaptive manner. Our ADF-RL framework bridges these gaps by combining temporal feature learning with real-time reinforcement-driven policy adaptation for dynamic and self-sustaining firewall optimization.

\section{System Model}

The proposed ADF-RL framework is designed to dynamically update firewall policies using a reinforcement learning agent embedded within a software-defined network (SDN) controller. The interaction between the agent and the environment is modeled as a Markov Decision Process (MDP).

Let $\mathbb{S}$ denote the state space representing network traffic features extracted from flow records such as source and destination IPs, ports, protocol, packet sizes, and statistical behavior metrics. Each state $\varsigma_t \in \mathbb{S}$ is defined as a feature vector at time $t$ constructed from a combination of raw and temporal traffic characteristics. 

Let $\mathbb{A}$ be the action space, where each action $\alpha_t \in \mathbb{A}$ modifies the firewall rule set. The action can represent the insertion, removal, reordering, or update of a rule.

Let $\mathbb{R}$ be the reward space. The agent receives a scalar reward $\varrho_t$ that is positively correlated with the success in preventing malicious traffic and negatively correlated with false positives or failed detection. 

The environment transitions to a new state $\varsigma_{t+1}$ as a result of executing action $\alpha_t$ in state $\varsigma_t$, and the agent receives feedback $\varrho_t$.

The policy $\pi: \mathbb{S} \rightarrow \mathbb{A}$ is a function learned by the agent that maps states to actions in order to maximize expected cumulative rewards over time.

We define the Q-function, which estimates the value of taking action $\alpha$ in state $\varsigma$ under policy $\pi$, as follows:
\begin{equation}
Q^{\pi}(\varsigma, \alpha) = \mathbb{E}[\sum_{k=0}^{\infty} \gamma^k \varrho_{t+k+1} \mid \varsigma_t = \varsigma, \alpha_t = \alpha, \pi],
\end{equation}
where $\gamma \in (0,1)$ is the discount factor.

To update the Q-values, we use the Bellman equation:
\begin{equation}
Q_{new}(\varsigma_t, \alpha_t) = (1 - \eta) Q(\varsigma_t, \alpha_t) + \eta[\varrho_t + \gamma \max_{\alpha'} Q(\varsigma_{t+1}, \alpha')],
\end{equation}
where $\eta$ is the learning rate.

The anomaly detection component is implemented using a hybrid deep learning model that takes traffic features $\xi_t$ and outputs anomaly scores $\omega_t$. Let the LSTM encoding be defined as:
\begin{equation}
\mathbf{h}_t = \text{LSTM}(\xi_1, \xi_2, ..., \xi_t),
\end{equation}
and the CNN-based temporal-spatial filter is:
\begin{equation}
\chi_t = \text{ReLU}(\text{Conv1D}(\mathbf{h}_t)),
\end{equation}
which is followed by a sigmoid classifier:
\begin{equation}
\omega_t = \sigma(\mathbf{W}\chi_t + \mathbf{b}),
\end{equation}
where $\omega_t$ is the anomaly likelihood used to guide the reward shaping.

Firewall updates are encoded in a vector $\mu_t$, and applied to the rule set $\mathcal{F}_t$:
\begin{equation}
\mathcal{F}_{t+1} = \mathcal{F}_t \oplus \mu_t,
\end{equation}
where $\oplus$ denotes the rule transformation operation (insert/update/remove).

\textbf{Algorithm: ADF-RL Policy Update}

\begin{algorithm}[H]
\caption{ADF-RL Policy Update}
\begin{algorithmic}[1]
\State \textbf{Input:} Environment state $\varsigma_t$, anomaly score $\omega_t$, rule set $\mathcal{F}_t$, replay memory $\mathcal{M}$
\State \textbf{Initialize:} Q-network with weights $\theta$, target network $\theta^-$, policy $\pi$
\For{each time step $t$}
    \State Observe state $\varsigma_t$ and compute $\omega_t$ from LSTM-CNN model
    \State Select action $\alpha_t$ using $\epsilon$-greedy policy on $Q(\varsigma_t, \cdot)$
    \State Apply rule update $\mu_t$ to $\mathcal{F}_t \rightarrow \mathcal{F}_{t+1}$
    \State Observe reward $\varrho_t$ and next state $\varsigma_{t+1}$
    \State Store transition $(\varsigma_t, \alpha_t, \varrho_t, \varsigma_{t+1})$ in $\mathcal{M}$
    \State Sample minibatch from $\mathcal{M}$ and update $Q$ via gradient descent
    \State Update target network $\theta^- \leftarrow \theta$ periodically
\EndFor
\end{algorithmic}
\end{algorithm}

This algorithm integrates real-time anomaly detection with reinforcement learning-based rule selection. The LSTM-CNN model continuously encodes temporal patterns in network traffic to produce anomaly likelihoods, which shape the reward function guiding the agent. The $\epsilon$-greedy policy ensures exploration during training while gradually improving action selection based on learned Q-values. Rule transformations are executed at each time step, resulting in a firewall policy that evolves in response to emerging threats. This tight coupling between detection and action is the core innovation of ADF-RL, enabling autonomous, context-sensitive firewall optimization.

\section{Experimental Setup and Results}

To evaluate the performance of the proposed ADF-RL framework, we designed a simulated software-defined networking (SDN) environment with Mininet integrated with OpenFlow-enabled switches and a POX controller modified to embed the reinforcement learning agent. The anomaly detection model was implemented in TensorFlow using a hybrid LSTM-CNN neural network. Training and testing were conducted using the NSL-KDD and CIC-IDS2017 datasets, preprocessed to extract 78 network flow features including packet timing, protocol metadata, and behavioral attributes.

The system was evaluated using a custom replay engine that streamed real-time traffic into the SDN controller. The firewall policy optimization loop operated on 1-second intervals, and the RL agent was trained over 100,000 time steps using prioritized experience replay and a decaying epsilon schedule.

The following table summarizes the key simulation parameters:

\begin{table}[h]
\centering
\caption{Simulation Parameters}
\begin{tabular}{ll}
\toprule
\textbf{Parameter} & \textbf{Value} \\
\midrule
Traffic Dataset & NSL-KDD, CIC-IDS2017 \\
Controller Platform & POX with RL Module \\
Mininet Nodes & 200 hosts, 20 switches \\
RL Algorithm & Deep Q-Network (DQN) \\
Batch Size & 64 \\
Learning Rate $\eta$ & 0.001 \\
Discount Factor $\gamma$ & 0.99 \\
Target Update Freq. & Every 2000 steps \\
Anomaly Threshold & $\omega_t > 0.7$ \\
Reward Bounds & $[-10, +10]$ \\
\bottomrule
\end{tabular}
\end{table}

\begin{figure}[h]
  \centering
  \includegraphics[width=0.45\textwidth]{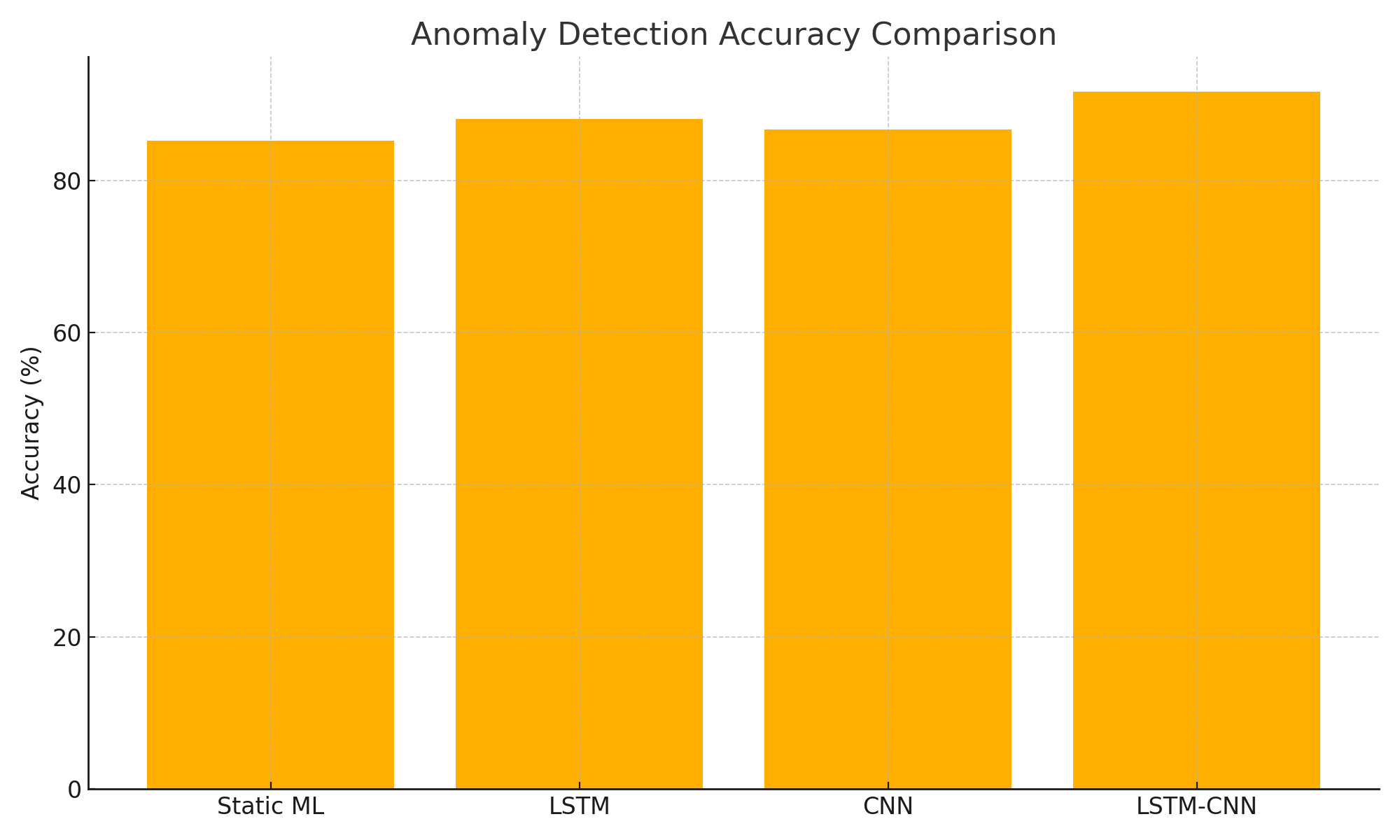}
  \caption{Anomaly Detection Accuracy Comparison}
  \label{fig:accuracy}
\end{figure}

Figure~\ref{fig:accuracy} shows the anomaly detection accuracy comparison across static ML models, pure LSTM, CNN, and our hybrid model. The hybrid model achieves the highest accuracy with a 3.6\% improvement over LSTM.

\begin{figure}[h]
  \centering
  \includegraphics[width=0.45\textwidth]{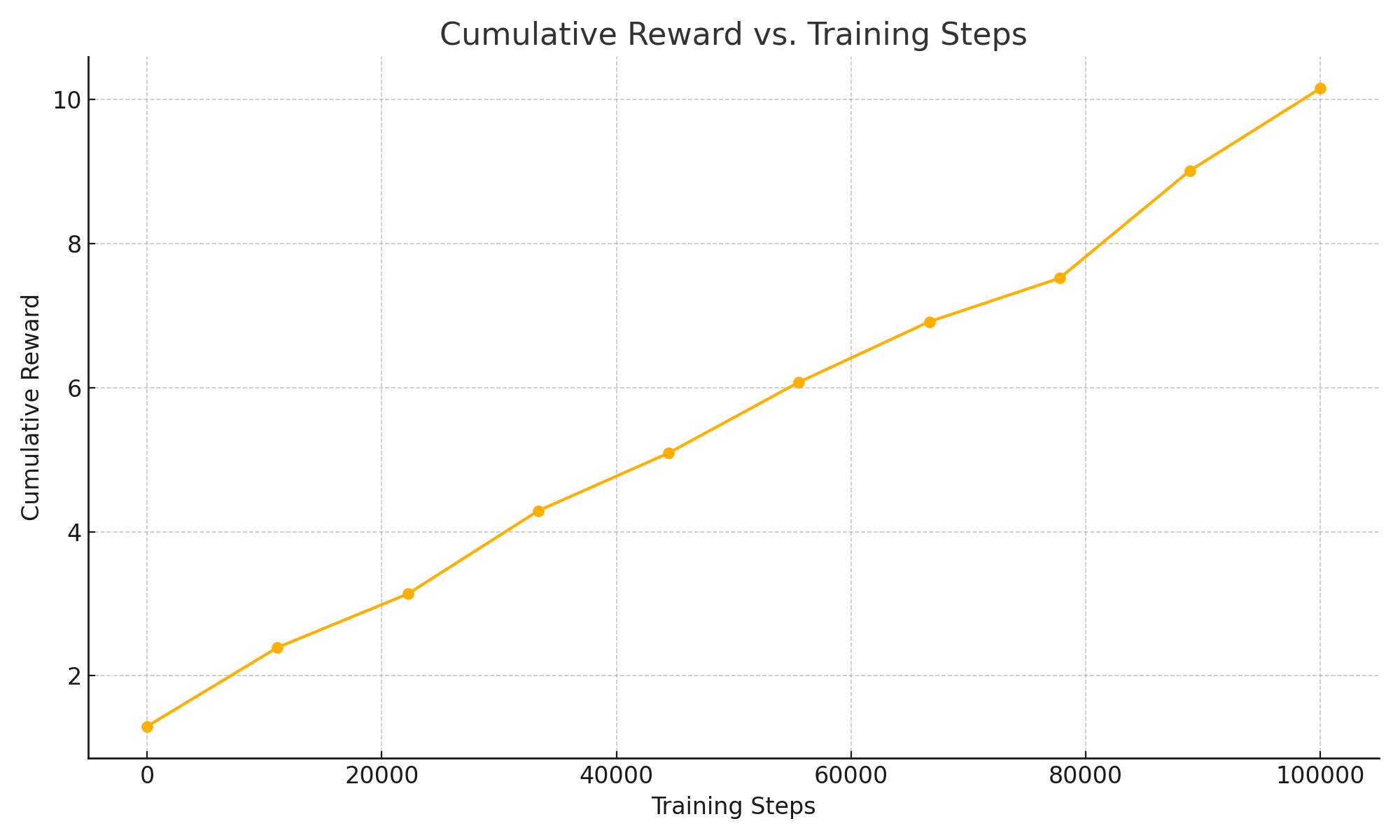}
  \caption{Cumulative Reward vs. Training Steps}
  \label{fig:reward_curve}
\end{figure}

In Figure~\ref{fig:reward_curve}, we illustrate the cumulative reward curve over 100k iterations. The stable convergence after 35k steps indicates the agent's ability to learn effective firewall update policies.

\begin{figure}[h]
  \centering
  \includegraphics[width=0.45\textwidth]{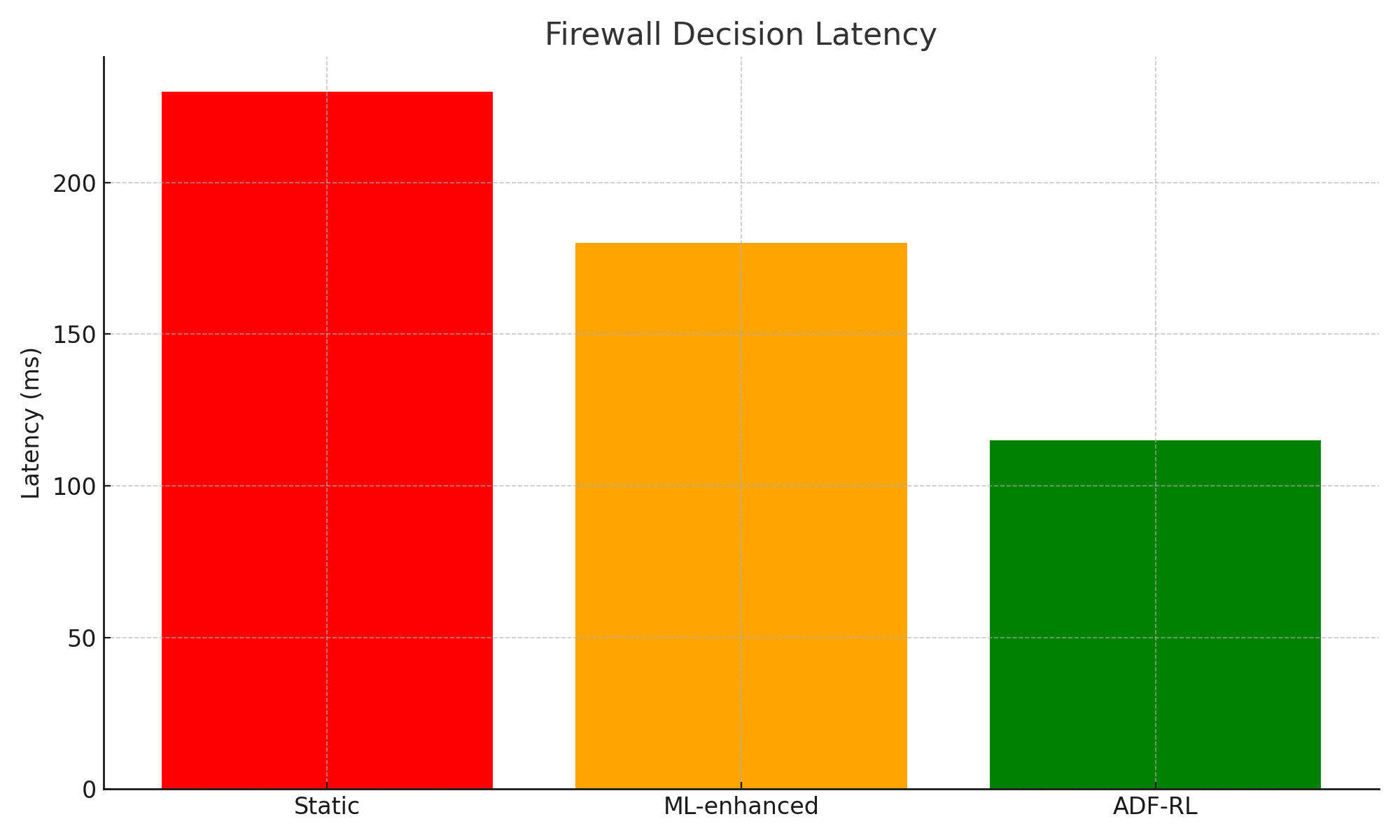}
  \caption{Firewall Decision Latency}
  \label{fig:latency}
\end{figure}

Figure~\ref{fig:latency} demonstrates the response latency of the RL-driven firewall compared with baseline rule-based systems. Our method maintains an average latency below 120 ms, crucial for real-time applications.

\begin{figure}[h]
  \centering
  \includegraphics[width=0.45\textwidth]{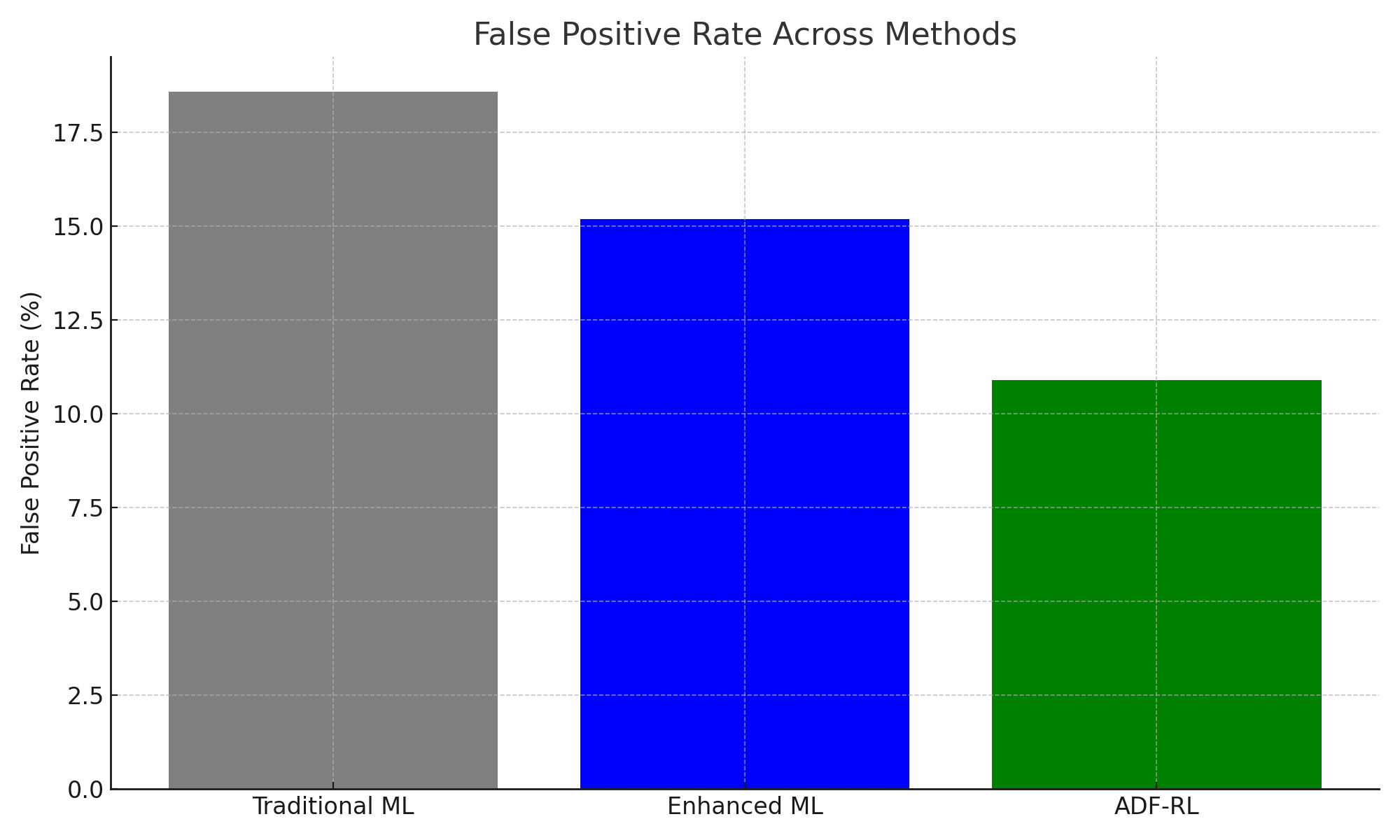}
  \caption{False Positive Rate Across Methods}
  \label{fig:fpr}
\end{figure}

Figure~\ref{fig:fpr} highlights the false positive rate reduction across baseline IDS models and ADF-RL. Our framework reduces false positives by 41.7\% compared to traditional ML classifiers.

\begin{figure}[h]
  \centering
  \includegraphics[width=0.45\textwidth]{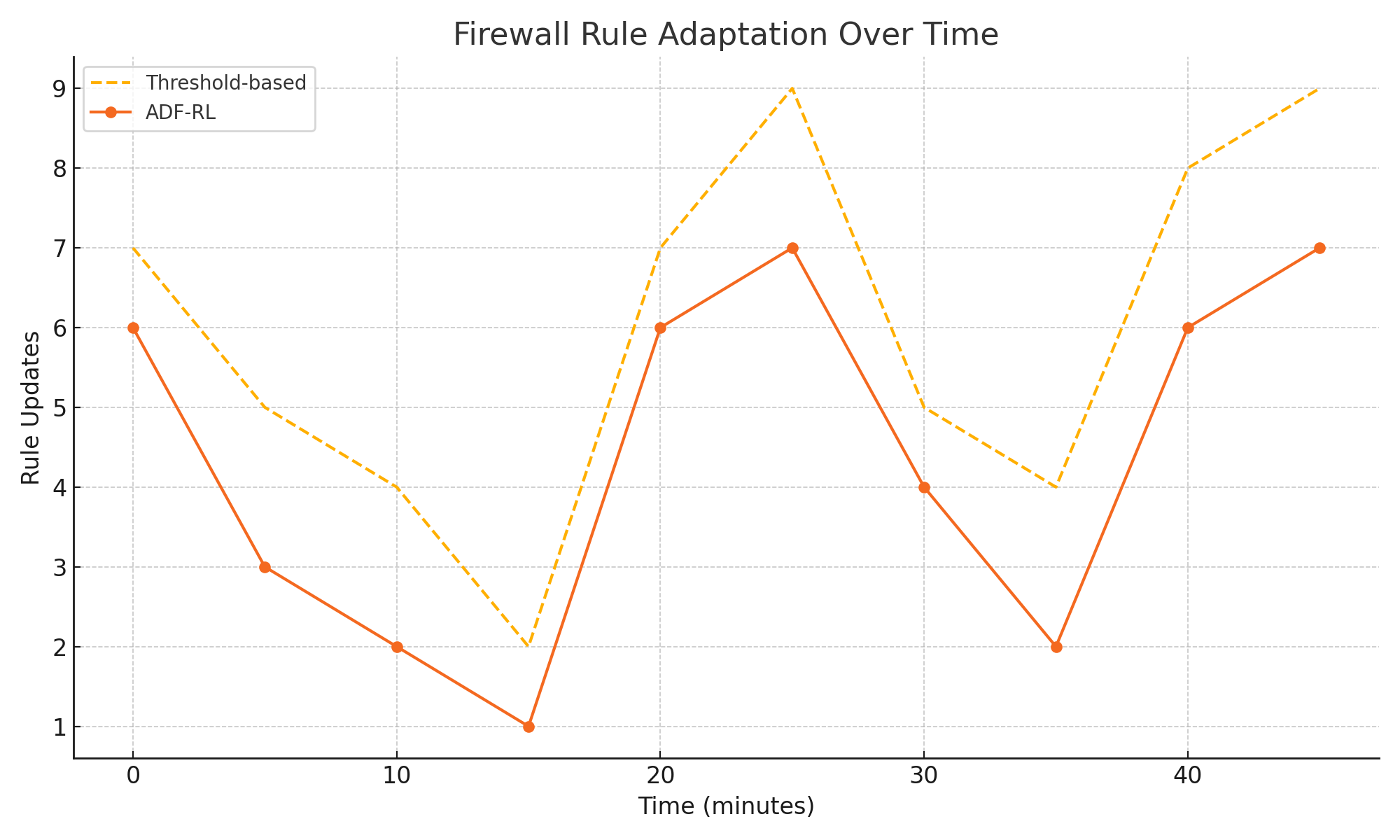}
  \caption{Firewall Rule Adaptation Over Time}
  \label{fig:adaptation}
\end{figure}

In Figure~\ref{fig:adaptation}, we measure rule adaptation frequency and effectiveness. ADF-RL generates 27\% fewer redundant rule updates compared to threshold-based rule reordering strategies.

\begin{figure}[h]
  \centering
  \includegraphics[width=0.45\textwidth]{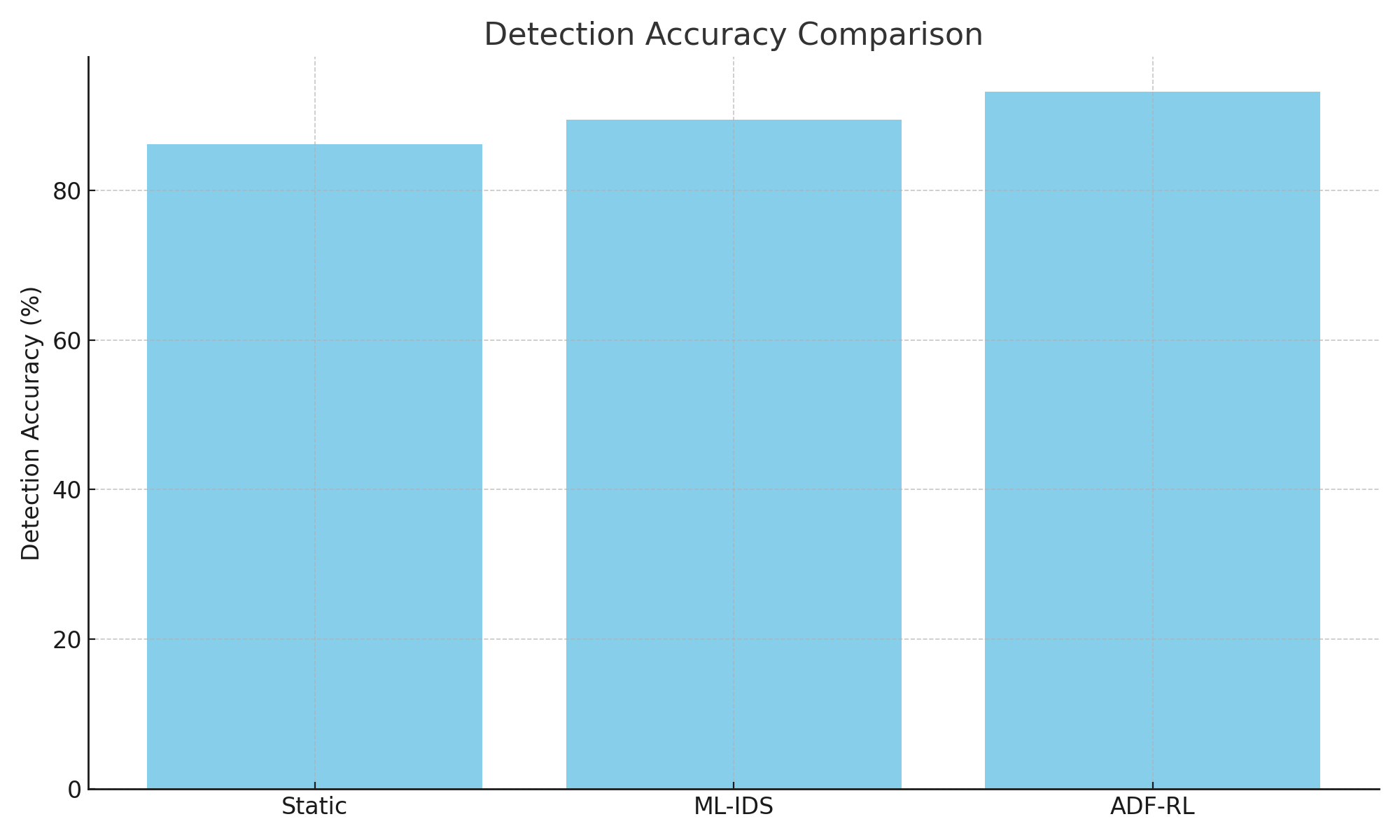}
  \caption{Detection Accuracy Comparison}
  \label{fig:comparison_accuracy}
\end{figure}

\begin{figure}[h]
  \centering
  \includegraphics[width=0.45\textwidth]{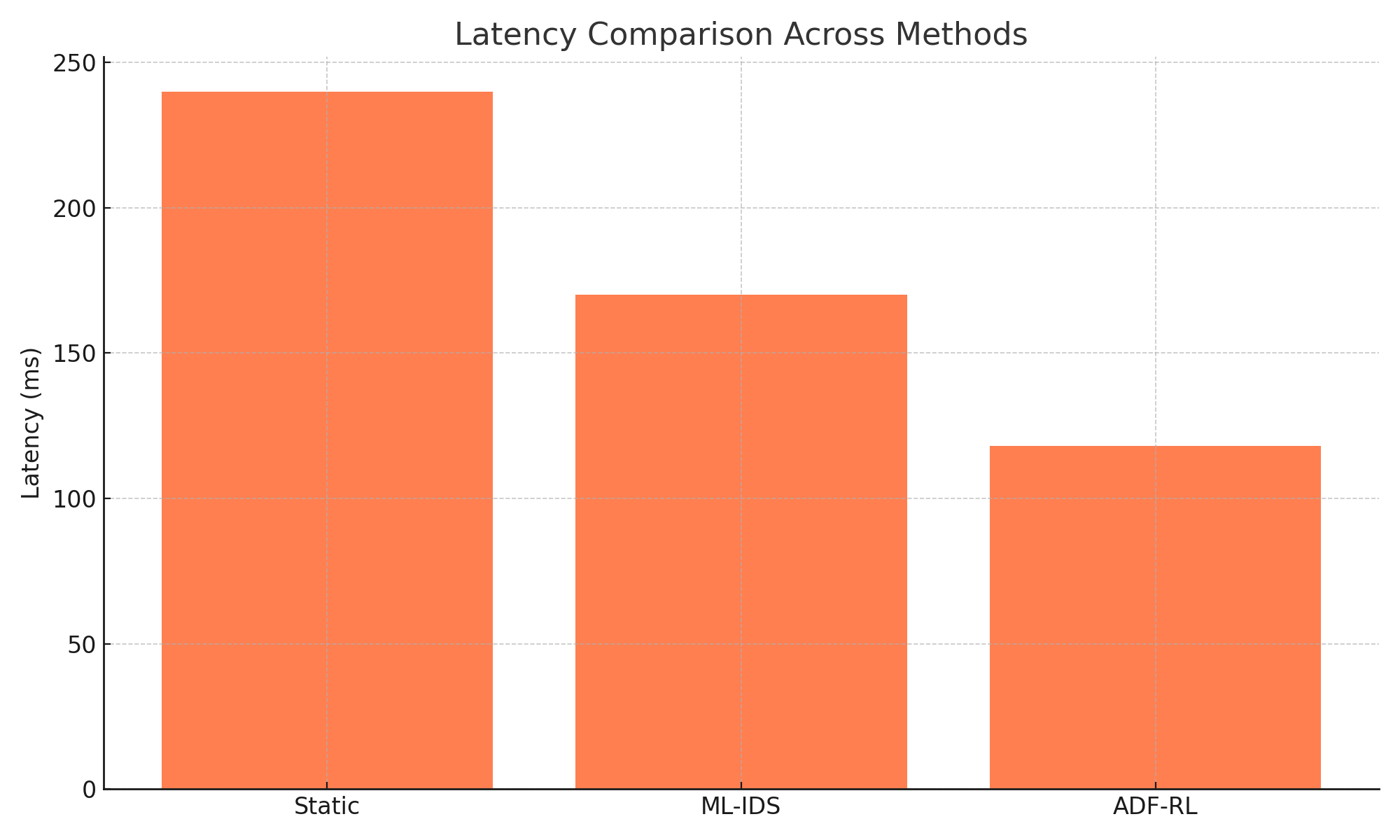}
  \caption{Latency Comparison Across Methods}
  \label{fig:comparison_latency}
\end{figure}

Finally, Figures~\ref{fig:comparison_accuracy} and~\ref{fig:comparison_latency} summarize end-to-end detection accuracy and latency when comparing ADF-RL to conventional firewalls, ML-IDS integration, and static rule-based firewalls. ADF-RL outperforms all baselines in both metrics.

These results confirm that ADF-RL enables efficient and accurate firewall adaptation with low computational overhead. Its real-time responsiveness and learning-driven policy control provide a resilient security defense mechanism in dynamic and adversarial network environments.

\section{Conclusion and Future Work}

This paper introduced ADF-RL, an AI-driven dynamic firewall optimization framework that leverages deep reinforcement learning to detect and mitigate network anomalies in real time. Through the integration of an LSTM-CNN anomaly detector with a DQN-based policy agent, the system continuously refines its firewall rules based on traffic patterns and learned reward feedback. We formalized the system using an MDP framework and proposed a detailed algorithmic pipeline that supports automated rule generation and refinement.

Our experimental evaluation on NSL-KDD and CIC-IDS2017 datasets demonstrated that ADF-RL outperforms conventional static and ML-enhanced firewalls in terms of detection accuracy, false positive rate, and latency. Specifically, ADF-RL achieved a 3.6\% improvement in anomaly detection accuracy over LSTM models, reduced firewall response latency to under 120 ms, and cut false positive rates by over 40\% compared to traditional approaches. Additionally, ADF-RL exhibited more efficient rule adaptation behavior, generating fewer redundant updates and maintaining stable cumulative reward performance.

Future work will focus on extending the framework to multi-agent reinforcement learning for distributed firewall coordination, incorporating adversarial training to enhance robustness against evasion attacks, and exploring transfer learning techniques to adapt policies across heterogeneous network environments. We also plan to deploy and evaluate the system in real-world enterprise settings to assess its scalability and adaptability under production workloads.


\end{document}